\documentclass[12pt]{article}
\usepackage{amssymb,amsmath,epsfig,mathtools,caption}
\usepackage{graphicx}

\begin{document}

\title{\bf Accreting Fluids onto Regular Black Holes Via Hamiltonian Approach}
\author{Abdul Jawad$^1$ \thanks{jawadab181@yahoo.com; abduljawad@ciitlahore.edu.pk}
and M.Umair Shahzad$^{1,2}$ \thanks{m.u.shahzad@ucp.edu.pk} \\
\small $^1$Department of Mathematics, COMSATS Institute of Information\\
\small Technology, Lahore-54000, Pakistan.\\
\small $^2$CAMS, UCP Business School, University of Central Punjab,\\
\small  Lahore, Pakistan}

\date{}
\maketitle

\begin{abstract}

We investigate the accretion of test fluids onto regular black holes
such as Kehagias-Sfetsos black hole and a regular black hole with
Dagum Distribution Function. We analyze the accretion process when
different test fluids are falling onto these regular black holes.
The accreting fluid is being classified through equation of state
according to features of regular black holes. The behavior of fluid
flow and the existence of sonic points is being checked for these
regular black holes. It is noted that three velocity depends on
critical points and equation of state parameter on phase space.

\end{abstract}

\section{Introduction}

A very interesting physical phenomena in astrophysics is the
accretion of fluids onto a black hole (BH), which has been
rigorously discussed in the literature. The presence of event
horizon is distinctive feature of BH which act as one way membrane
through which accreting fluid or gas disappears and leads to several
important issues. For instance, radiations emitted near the BH
undergo the effect of strong gravitational lensing and appears as
the image of BH's shadow surrounded by sharp light ring. However, no
inner boundary conditions are necessary for equation of motion in
case of BH accretion \cite{2}. On the other hand, one of the major
challenge in general relativity is the existence of essential
singularities and it looks like the common property in most of
solutions of Einstein field equations.

Moreover, regular black holes (RBHs) have been constructed to
resolve this problem because their metrics are regular everywhere
and hence essential singularities can be avoided in the solutions of
Einstein equations of BHs physics \cite{c28}. These RBHs satisfied
the weak energy condition while some of them violate the strong
energy conditions \cite{c17,c18}. However, Penrose cosmic censorship
conjecture claims that singularities predicted by general relativity
occur and \textbf{they must be enclosed by event horizon}
\cite{c19,c20}. In this way, Bardeen \cite{c29} made the pioneer
work and obtained a BH solution without any essential singularity at
origin enclosed by event horizon known as 'Bardeen Black Hole' which
satisfy weak energy condition. Later, many authors found similar
solution \cite{c30,c31,c32}. The coupling of general relativity to
non-linear electromagnetic theory has brought a new sets of charged
RBHs. Hayward \cite{c34} and Berej et. al \cite{c33} found different
kinds of RBH solutions. Recently, Leonardo et. al \cite{c22} used
various distribution functions (such as logistic, Fermi-Dirac, Dagum
etc.) in order to obtain charged RBH. Kehagias-Sftesos (KS) found
asymptotically flat RBH which usually behaves like Schwarzschild
(SH) BH in Horava theory \cite{28}.

The pioneer work on accretion for spherical symmetry was introduced
by Bondi in the frame work of Newtonian gravity \cite{8}. Many
authors have investigated Bondi-type accretion flows in SH, SH
de-Sitter and SH anti-de-Sitter BHs \cite{12,13,23}. In this
extension, Michel \cite{9} proposed its general relativistic version
by considering the steady state flow of perfect fluid onto BH along
the radial direction. The stability of Michel type accretion in
subsonic region has been analyzed by Moncrief \cite{10}. Further, it
was suggested by Babichev et al. \cite{17} that BH loses its mass
during the accretion of phantom dark energy onto it. Furthermore,
many authors have worked on radial flows such as accretion of dark
matter onto BH \cite{11}, radial accretion onto cosmological BHs
\cite{12,13}, for self gravitating \cite{14,15} and perturbation
theories \cite{16}. The accretion onto Einstein - Maxwell - Gauss -
Bonnet BH and topologically charged BHs in $f(R)$ theories has been
presented in \cite{21,22}. The nonlinear study of phantom scalar
field accreted onto a BH was done Gonzalez and Guzman \cite{i1} with
the help of numerical methods. In this direction, some works have
been done by various people \cite{18,19,c28} and showed that when
phantom-like fluids accrete onto BHs/RBHs then mass of BHs/RBHs
decrease, respectively.

Recently, Chaverra et al. \cite{2} and Ahmed et al. \cite{24,25}
have discussed Michel-type accretion near SH, $f(R)$ and $f(T)$ BHs.
In the present work, we will use similar technique in order to
discuss the accretion onto well-known RBHs. Rest of the paper is
organized as follows: In section \textbf{2}, we derive general
formalism for spherically static accretion process. In section
\textbf{3}, we discuss the accretion process using Hamiltonian
dynamical system and study the system at sonic points. In section
\textbf{4}, we study two RBHs: KS-BH and RBH using Dagum
Distribution Function (DDF RBH). In section \textbf{5}, we find the
solutions for isothermal test fluids of each RBH for different kinds
of fluids. In section \textbf{6}, we investigate the accretion
phenomena for polytropic fluid. In the last section, we summarize
our results.


\section{General equations of spherical accretion}

Here, we will derive the governing equations and analyze the flow of
perfect fluid and accretion rate onto RBHs by utilizing the energy
and particle conservation laws. In this way, we consider general
spherical symmetric and stationary line element as follows
\begin{equation}\label{a1}
ds^{2}=-X(r)dt^{2}+\frac{1}{X(r)}dr^{2}
+r^2(d\theta^{2}+\sin^{2}\theta d\phi^{2}),
\end{equation}
where $X(r)>0$ is the function of $r$. The energy momentum tensor
for perfect fluid is given by
\begin{equation}\label{a2}
T_{\mu\nu} = (p + \varepsilon)u_\mu u_\nu + p g_{\mu\nu},
\end{equation}
where $p$ is the pressure, $\varepsilon$ is the energy density and
$u^{\mu}$ is the four velocity which is given by
\begin{equation}\label{3}
u^{\mu} = \frac{dx^{\mu}}{d\tau} = (u^{t},u^{r},0,0),
\end{equation}
where $\tau$ is the proper time. Here, $u^{\theta}$ and $u^{\phi}$
become zero due to spherical symmetry restrictions. We define the
current density or particle flux by $J^{\mu}=n u^{\mu}$. According
to law of particle conservation, the divergence of particle flux is
zero for this system, i.e.,
\begin{equation}\label{a3}
\nabla_{\mu}J^{\mu}=\nabla_{\mu}(n u^{\mu})=0,
\end{equation}
where $\nabla_{\mu}$ is covariant derivative. It is useful to
mention here that all the flow variables $(p, \varepsilon, u^{\mu},
n)$ are spherically symmetric and steady state according to
definition of Michel \cite{12,13}. Using Eq.(\ref{a1}) and
normalization condition $u^{\mu}u_{\mu}=-1$, we have
\begin{equation}\label{a4}
u_t=\pm \sqrt{u^2+X(r)}.
\end{equation}
where $u=dr/d\tau=u^{r}$, $u^{t}$ can be negative or positive due to
square root which represents the backward or forward in time
conditions. However, $u < 0$ is required for accretion process
otherwise for any outward flows $u > 0$.
The equation of
continuity is
\begin{equation}\label{a5}
\nabla_{\mu}(n u^{\mu}) = \frac{1}{r^2} \partial_{r}(r^2 n u)=0,
\end{equation}
which leads to
\begin{equation}\label{a6}
r^2nu=C_a,
\end{equation}
where $C_a$ is integration constant. The thermodynamics of perfect
fluid is given by \cite{26}
\begin{equation}\label{a7}
dp=n(dh-Tds), ~~~ d\varepsilon=hdn+nTds,
\end{equation}
where $h=\frac{\varepsilon+p}{n}$ is the specific enthalpy, \emph{T}
is the temperature and \emph{s} is the specific entropy.

Furthermore, the theorem of relativistic hydrodynamics (which is
only applicable for smooth flows) states that the scalar quantities
such as $h u _{\mu} \eta^{\mu}$ remain conserved along the
trajectories of the fluid \cite{26}, i.e.,
\begin{equation}\label{a8}
u^{\nu}\partial_{\mu}(h u _{\mu} \eta^{\mu})=0,
\end{equation}
where $\eta^{\mu}$ is a Killing vector of space-time. Considering
$\eta^{\mu}=(1, 0, 0, 0)$ of metric (\ref{a1}), we obtain
\begin{equation}\label{a9}
h\sqrt{u^2+X(r)}=C_b
\end{equation}
where $C_b$ is integration constant. It is also mention here that
the specific entropy remains conserved along the fluid trajectories,
i.e., $u^{\mu}\nabla_{\mu}s =0$. Hence, the stress energy tensor
$T_{\mu \nu}$ can be written as \cite{24,25}
\begin{equation}\nonumber
T_{\mu\nu} = nh u_\mu u_\nu + (nh-\varepsilon) g_{\mu\nu}
\end{equation}
and then projection of $T^{\mu \nu}$ onto $u^{\mu}$ turns out to be
\begin{equation}\label{a10}
u_{\nu}\nabla_{\mu}T^{\mu \nu}=u_{\nu}\nabla_{\mu}[nh u_\mu u_\nu +
(nh-\varepsilon) g_{\mu\nu}] = -nTu^{\mu}\nabla_{\mu}s=0.
\end{equation}
Since fluid is stationary (independent of time) and moving along
radial direction only in the present case which leads to $\partial_r
s=0$ and hence $s$ becomes constant. In this scenario, Eq.
(\ref{a7}) reduces to
\begin{equation}\label{a11}
dp = n dh, ~~~ d\varepsilon = h dn.
\end{equation}
Next, we will use Eqs.(\ref{a6}), (\ref{a9}) and (\ref{a11}) to
analyze the flow. Since s becomes constant and thus canonical form
of EoS of simple fluid $(\varepsilon = \varepsilon(n, s))$
\cite{24,25} turns out to be barotropic form and is given by
\begin{equation}\label{a12}
\varepsilon = F(n).
\end{equation}
In view of above relation, Eq.(\ref{a11}) leads to
\begin{equation}\label{a13}
h = \frac{d\varepsilon}{dn} = \frac{dF(n)}{dn} = F'(n),\quad p'=nF''
\Longrightarrow p = nF'-F,
\end{equation}
which provides the relationship between \emph{F} and \emph{G} $(p =
G(n))$as follows
\begin{equation}\label{a16}
G(n)=nF'(n)-F(n).
\end{equation}
Further, the sound speed can be defined as $a^2=(\partial p/\partial
\varepsilon)_s$, which reduces to $a^2=dp/d\varepsilon$ because $s$
is constant. Using (\ref{a11}), we can find
\begin{equation}\label{a17}
\frac{dh}{h}=a^2 \frac{dn}{n}.
\end{equation}
Using Eqs.(\ref{a13}) and (\ref{a17}), we obtain
\begin{equation}\label{a18}
    a^2 = n(\ln F')'.
\end{equation}
Since the motion is in radial direction, so the metric (\ref{a1})
reduces to
\begin{equation}\label{a19}
    ds^2 = -(\sqrt{X(r)} dt)^2+\bigg(\frac{dr}{\sqrt{X(r)}}\bigg)^2.
\end{equation}
The ordinary three velocity of the fluid can be defined as
$v\equiv\frac{dr/\sqrt{X(r)}}{\sqrt{X(r)}dt}$ and its expression can
be obtained by using relations $u^t=dt/d\tau$, $u=u^r=dr/d\tau$,
$u_t=-X(r)u^t$ and Eq.(\ref{a4}) as follows
\begin{equation}\label{a20}
    v^2=\frac{u^2}{X(r)+u^2},
\end{equation}
which implies
\begin{equation}\label{a21}
u^2=\frac{X(r) v^2}{1-v^2} ~~ and ~~ (u_t)^2=\frac{X^2(r)}{1-v^2}.
\end{equation}
Utilizing above relations in (\ref{a6}), we obtain
\begin{equation}\label{a22}
    C_a^2=\frac{r^4n^2X(r)v^2}{1-v^2}.
\end{equation}
These results will be used in the following sections \cite{24,25}.

\section{Hamiltonian system}

Here, we derive two integrals of motion $(C_a, C_b)$ given in
(\ref{a6}) and (\ref{a9}) by adopting Hamiltonian procedure. The
idea of reformulating of the Michel flow problem as a Hamiltonian
system for global flow is analyzed in detail by
\cite{2,27a,27b,27c}. The Hamiltonian $H$ is a function of two
variables $(x, y)$ and its simplest form has one degree of freedom.
Let $H$ be the square of L.H.S of Eq.(\ref{a9}):
\begin{equation}\label{h1}
    H=h^2(X(r)+u^2).
\end{equation}
Using Eq.(\ref{a21}) in above equation, we have
\begin{equation}\label{h2}
    H(r,v)=\frac{h(r,v)^2X(r)}{1-v^2},
\end{equation}
and this is derived in \cite{24,25}. Here, we fixed dynamical
variables to be $(r,v)$. For the derivation of critical points,
particularly, the sonic points are derived in following subsection.

\subsection{Sonic points}

The dynamical system of Hamiltonian $H$ given in (\ref{h2}) reads to
be \cite{24,25}
\begin{equation}\label{h3}
\dot{v}=-H_{,~r}, ~ \dot{r}=H_{,~v},
\end{equation}
where dot represents the derivative with respect to $\bar{t}$ which
is new \textbf{"time"} variable. The \textbf{"time"} variable
$\bar{t}$ for the dynamical system is any variable on which Eq.
(\ref{h1}) does not depend explicitly so that the dynamical system
is autonomous. For finding critical points (CPs), we utilize
Eq.(\ref{h3}) for which we require
\begin{eqnarray}
  \frac{d}{dr}H(r,v) &=& \frac{h^2}{1-v^2}\left(\frac{d}{dr}(X(r))+2X(r)\frac{d}{dr}(\ln h)\right),
  \\\label{h4}
  \frac{d}{dv}H(r,v) &=& \frac{2X(r) h^2 v}{(1-v^2)^2}\left(1+\frac{1-v^2}{v^2}\frac{d}{dv}(\ln
  h)\right).\label{h5}
\end{eqnarray}
Moreover, Eq.(\ref{a17}) can be expressed as
\begin{eqnarray}
\frac{d}{dr}(\ln h)&=& a^2 \frac{d}{dr}(\ln n),\\\label{cp1}
\frac{d}{dv}(\ln h)&=& a^2 \frac{d}{dv}(\ln n).\label{cp2}
\end{eqnarray}
By keeping '$r$' as a constant in Eq.(\ref{a22}), we have $\frac{n
v}{\sqrt{1-v^2}} = \text{constant}$ which under differentiation
w.r.t. '$v$' leads to
\begin{equation}\label{cp3}
    \frac{d}{dv}(\ln n)=-\frac{1}{v (1-v^2)}\Rightarrow \frac{d}{dv}(\ln
    h)=-\frac{a^2}{v(1-v^2)}.
\end{equation}
Similarly,
\begin{equation}\label{cp3}
    \frac{d}{dr}(\ln n)=-\frac{4+ r \frac{d}{dr}(\ln X(r))}{2r}\Rightarrow \frac{d}{dr}(\ln
    h)=-\frac{a^2\big(4+ r \frac{d}{dr}(\ln X(r))\big)}{2r}.
\end{equation}
Finally, by using Eqs.(\ref{h3})-(\ref{cp3}), we obtain
\begin{eqnarray}
  \dot{v} &=& -\frac{h^2}{r(1-v^2)}\left(r \big(\frac{d}{dr}X(r)\big)(1-a^2)-4 X(r) a^2\right),
  \\\label{h6}
  \dot{r} &=& -\frac{2 X(r)
  h^2}{v(1-v^2)^2}\left(v^2-a^2\right).\label{h7}
\end{eqnarray}
By setting the above relations equal to zero, we can get CPs as
follows
\begin{equation}\label{h8}
    v_c^2=a_c^2 ~~and~~ r_c(1-a_c^2)\frac{d}{dr_c}X(r_c) = 4 X_c a_c^2.
\end{equation}
Here $X_c\equiv X(r_c)$ and $\frac{d}{dr_c}X_{c}\equiv
\frac{d}{dr}X(r)\mid_{~r=r_c}$. The second relation of Eq.(\ref{h8})
represents the sound speed at the CP, i.e., $a_c^2$ in terms of
$r_c$
\begin{equation}\label{h9}
a_c^2=\frac{r_c \frac{d}{dr_c}X_{c}}{4 X_c + r_c
\frac{d}{dr_c}X_{c}},
\end{equation}
In this scenario, the constant $C_a^2$ (\ref{a22}) can be written as
\begin{equation}\label{h10}
C_a^2=r_c^4 n_c^2 v_c^2\frac{X_c}{1-v_c^2} = \frac{1}{4}r_c^5 n_c^2
\frac{d}{dr_c}X_{c},
\end{equation}
where we have used (\ref{h8}). Using $C_a^2$ in Eq.(\ref{a22}), we
have
\begin{equation}\label{h11}
\left(\frac{n}{n_c}\right)^2=\frac{r_c^5
\big(\frac{d}{dr_c}X_{c}\big)(1-v^2)}{4r^4 X(r)v^2}.
\end{equation}

\section{Regular Black holes}

In this section, we will discuss RBHs with fixed BH background,
neglecting the self-gravity of the fluid.

\subsection{KS Regular Black hole}

Recently, Horava proposed the renormalizable gravity theory with
higher order spatial derivatives in four dimensional space-time.
\textbf{It is an ultraviolet completion of general relativity. This
theory reduces to Einstein gravity with non-vanishing cosmological
constant under infrared limit, but it has improved ultraviolet
behaviors.} Moreover, in deformed Horava-Lifshitz (HL) gravity, the
ultraviolet properties are unchanged, whereas there exists a
Minkowski vacuum in infra-red limit. The HL theory is considered as
very interesting and many researchers discovered new BH solutions
after its formulation \cite{27,28,29,30}. Many new aspects discussed
in connection with HL theory \cite{31}. The metric function of KS
RBH in deformed HL gravity is given by
\begin{equation}\label{ks2}
    X(r) = 1 + br^2\left(1-\sqrt{1+\frac{4M}{br^3}}\right),
\end{equation}
where $b$ is the arbitrary constant and $M$ is BH mass. However, the
constraints on the value of $bM^2$ have been developed through the
comparison of perihelion shift test of KS BH with the observations
in the solar system. It was found that $bM^2 \geq 1.7 \times
10^{-12}$ for Saturn, $bM^2 \geq 9 \times 10^{-12}$ for Mars and
$bM^2 \geq 7.2 \times 10^{-10}$ for Mercury \cite{32}. KS BH can be
reduced to $X(r)=1-\frac{2M}{r}+\frac{2M^2}{br^4}+...$ in the
limiting case $b\rightarrow \infty$ or $r\rightarrow \infty$, while
asymptotically behaves as the SH metric. The infra-red properties of
KS and RN BHs are different as shown in third term of expansion
\cite{33}. The KS metric have two horizons $r_h$ (outer) and
$r_{ch}$ (inner) and can be obtained from the metric function
(\ref{ks2}) as follows
\begin{equation}\label{1eeee}
r_{ch} = M\left(1- \sqrt{\left(1-\frac{1}{2 b M^2}\right)}\right) ~~
and ~~ r_{h} = M\left(1+ \sqrt{\left(1-\frac{1}{2 b
M^2}\right)}\right),
\end{equation}
with $2bM^2\geq1$ \cite{34}. For $2bM^2 = 1$, we have an extreme BH
and for $2bM^2 < 1$, we have naked singularity.

\subsection{Regular Black hole using Dagum Distribution Function}

Consider the line element (\ref{a1}) for generally spherically
symmetric metric with
\begin{equation}\label{rv1}
    X(r) = 1-\frac{2M}{r} \frac{\xi (r)}{\xi (\infty)},
\end{equation}
where the Dagum distribution function \cite{35a} is
\begin{equation}\label{rv2}
    \xi(x)=\frac{ap x^{ap-1}}{b^{ap}
    \Big(1+\big(\frac{x}{b}\big)^a\Big)^{p+1}},
\end{equation}
and $a,b,p$ are positive parameters. By assuming $b=1$ and $p=1/a$,
we find
\begin{equation}\label{rv3}
    \xi(x)=\frac{1}{\Big(1+x^a\Big)^{\frac{a+1}{a}}}.
\end{equation}
After replacing $x\rightarrow \frac{q^2}{Mr}$ and simplifications
leads to
\begin{equation}\label{rv4}
    X(r) = 1-\frac{2M}{r}\left(\frac{1}{\big(1+r
    \big(\frac{q^2}{Mr}\big)^a\big)^{\frac{a+1}{a}}}\right)^{\beta}.
\end{equation}
This BH satisfies the weak energy condition for $\beta =
\frac{3}{a+1}$ and hence we have
\begin{equation}\label{rv5}
X(r) = 1-\frac{2M}{r}\left(\frac{1}{1+r
\Big(\frac{q^2}{Mr}\Big)^a}\right)^{3/a}.
\end{equation}
Moreover, one can find the RBH metric by using Dagum distribution
function with a factor $q/r^2$ which behaves asymptotically as RN
metric and its metric function is
\begin{equation}\label{r2}
    X(r)=1-\frac{2M}{r}\left(\frac{1}{1+\gamma\left(\frac{q^2}{Mr}
    \right)^a}\right)^{\frac{3}{a}}+\frac{q^2}{r^2}\left(\frac{1}
    {1+\gamma\left(\frac{q^2}{Mr}\right)^a}\right)^{\frac{4}{a}},
\end{equation}
where $\gamma >0$ is a constant and $a\geq2$ is an integer. It
should be noted that the associated solution satisfies the weak
energy condition if we have $\gamma \geq (2/3)^a$ as shown in
\cite{35}. The associated electric field expression is given by
\begin{equation}
    E = \frac{q}{r^2}\left(\frac{3 \gamma (3+a)\left(\frac{q^2}{Mr}
    \right)^{a-1}}{2\left(1+\gamma \left(\frac{q^2}{Mr}\right)^a\right)
    ^{2+\frac{3}{a}}}+\frac{1-3 \gamma (3+a)\left(\frac{q^2}{Mr}
    \right)^{a}}{\left(1+\gamma \left(\frac{q^2}{Mr}\right)^a\right)
    ^{2(2+a)/a}}\right).
\end{equation}
The metric with function (\ref{r2}) behave-like de Sitter BH metric
as
\begin{equation}
    X(r)\approx 1-\frac{M^4}{\gamma^{\frac{4}{a}}q^6}(2
    \gamma^{\frac{1}{a}}-1)r^2.
\end{equation}
It can be noticed that \textbf{the term proportional to $r^2$}
cannot become zero because $\gamma \geq (2/3)^a$. The metric
function (\ref{r2}) remains regular, if we set $(1/2)^a\leq \gamma
<(2/3)^a$ but they have de Sitter center without satisfying the weak
energy condition. Therefore, if BH metric is regular and satisfies
the weak energy condition, then it has a de Sitter center \cite{36}.
However, if the metric has a de Sitter behavior when approaching to
the center, it does not necessarily satisfies the weak energy
condition. Furthermore, if we set $\gamma < (1/2)^a$, BH metric is
not regular. Finally, there is a known case which can be obtained as
a particular case of Eq.(\ref{r2}) by choosing $a = 2$ and $\gamma =
\frac{M^2}{q^2}$,i.e,
\begin{equation}\label{imp2}
    X(r)=
    1-\frac{2M\left(\frac{r^2}{r^2+q^2}\right)^{3/2}}{r}+
    \frac{q^2}{r^2}\left(\frac{r^2}{r^2+q^2}\right)^2.
\end{equation}
This case corresponds to the RBH metric given in \cite{37} and we
will discuss the accretion onto this RBH in the following sections.

\section{Isothermal test fluids}

The fluid flowing at a constant temperature is known as isothermal
fluids. In accretion process, sound speed of the fluid flow remains
constant. Moreover, the speed of sound at sonic points is equal to
the speed of sound of accretion flow at any radius \cite{24}. Here,
we follow \cite{25} to find the general solution of isothermal EoS
of the form $p=k \varepsilon $ which leads to $p=k F(n)$ and $G(n)=k
F(n)$ given in Eqs. (\ref{a12}) and (\ref{a16}), respectively, where
$k$ $(0<k\leq 1)$ is the state parameter. The differential equation
(\ref{a16}) becomes
\begin{equation}\label{is1}
nF'(n)-F(n) = k F(n),
\end{equation}
using (\ref{a12}) and integrating above equation, we have
\begin{equation}\label{is2}
    \varepsilon = F = \frac{\varepsilon_c}{n_c^{k +1}}n^{k +1},
\end{equation}
where $\frac{\varepsilon_c}{n_c^{k +1}}$ appears as integration
constant and hence Eq.(\ref{a13}) becomes
\begin{equation}\label{is3}
h=\frac{(k +1) \varepsilon_c}{n_c} \left(\frac{n}{n_c}\right)^{k}.
\end{equation}
Using Eq.(\ref{h11}), we have
\begin{equation}\label{is4}
h^2 \propto \left(\frac{1-v^2}{v^2 r^4 X(r)}\right)^{k},
\end{equation}
and
\begin{equation}\label{is5}
H(r,v) = \frac{X^{1-k}(r)}{\left(1-v^2\right)^{1-k}v^{2k} r^{4 k}},
\end{equation}
where all constants get absorb in the redefinition of Hamiltonian
$H$ and the time $\bar{t}$. Next, we need to discuss the pressure
physically. Using EoS $p=k \epsilon$ and by substituting
Eq.(\ref{h11}) in (\ref{is2}), we get
\begin{equation}\label{pp1}
p\propto \Big(\frac{1-v^2}{X(r)v^2 r^4}\Big)^{\frac{k+1}{2}}.
\end{equation}
If this pressure approaches to the event horizon from the region
where $t$ is time like, then $X(r)\rightarrow 0$ and the speed $v
\rightarrow 0 ~~\text{or}~~ 1$. In this situation, Hamiltonian
remains constant on the solution curve. For $v\rightarrow 0$, the
solution curve approaches to horizon and hence pressure diverges.
For $v \rightarrow 1$, Eq. (\ref{pp1}) remains finite in the
surrounding of horizon \cite{24,25}. If $X(r)=0$ has single root as
$r\rightarrow r_h$ then using (\ref{pp1}), one can observe the
pressure diverges as
\begin{equation}\label{pp2}
    p \sim \Big(r-r_h\Big)^{-\frac{k+1}{2k}}.
\end{equation}
If $X(r)=0$ has double root then
\begin{equation}\label{pp3}
    p \sim \Big(r-r_h\Big)^{-\frac{k+1}{k}}.
\end{equation}

The Hamiltonian (\ref{is5}) of the dynamical system is a constant
along the solution curve. A global flow solution that extends to
spatial infinity is
\begin{equation}\label{fl1}
    v\simeq v_1 r^{-\alpha}+v_{\infty} ~ as ~ r\rightarrow \infty,
\end{equation}
where $(v_1,|v_{\infty}|\leq 1, \alpha > 0)$ are constants.
Substituting Eq. (\ref{fl1}) in Hamiltonian (\ref{is5}), we obtain
\begin{eqnarray}
  H &\simeq& \frac{X^{1-k}(r)}{r^4k} ~~~~ 0< |v_{\infty}|<1, \label{fl2} \\
  H &\simeq& \frac{X^{1-k}(r)}{r^{(4-2\alpha)k}} ~~~~ |v_{\infty}|=0, \label{fl3} \\
  H &\simeq& \frac{X^{1-k}(r)}{r^{(4k+k \alpha-\alpha)}} ~~~~
  |v_{\infty}|=1 \label{fl4}.
\end{eqnarray}

Now the behavior of the fluid is analyzed by considering different
choices of state parameter $k$ such as $k = 1/4$ (sub-relativistic
fluid), $k = 1/3$ (radiation fluid), $k = 1/2$ (ultra-relativistic
fluid) and $k = 1$ (ultra-stiff fluid) and these are discussed in
following subsections.

\subsection{Solution for ultra-stiff fluid $(k = 1)$}

We assume the fluids with EoS $p=\epsilon$ and these are ultra-stiff
fluids. For instance, the usual EoS for the ultra-stiff fluids is $p
= k \varepsilon$, i.e., the value of state parameter is defined as
$k = 1$. The Hamiltonian (\ref{is5}) reduces to
\begin{equation}\label{u1}
    H=\frac{1}{v^2 r^4}.
\end{equation}
In this case, the metric function $X(r)$ does not involve and it is
discussed in detail in \cite{2,24,25}.

\subsection{Solution for ultra-relativistic fluid $(k=1/2)$ }

Here, we consider the fluids with EoS $p=\epsilon/2$ and these
fluids are ultra-relativistic. The Hamiltonian takes the simple form
\begin{equation}\label{s1}
H = \frac{\sqrt{X(r)}}{r^2 |v| \sqrt{1-v^2}}
\end{equation}
It is clear from this expression that the point $(r,v^2) = (r_h,1)$
is not a critical point of the dynamical system. For some given
value of $H=H_c$, Eq. (\ref{s1}) can be solved for $v^2$ as follows
\begin{equation}\label{2}
v^2=\frac{1\pm \sqrt{1-4f(r)}}{2}
\end{equation}
where $f(r)\equiv \frac{X(r)}{H_c r^4}$. We plot the above velocity
of fluid ($v$) versus $r$ for aforementioned two RBHs such as KS RBH
and DDF RBH by inserting the corresponding values of $X(r)$ as shown
in the left and right panels of Fig. \textbf{1}, respectively. The
trajectories in Fig. \textbf{1} corresponds to values of $H_0=\{H_c,
H_c\pm 0.04, H_c\pm0.09$\} where $H=H_c=0.170615$ (for KS RBH) and
$H=H_c=0.170615$ (for DDF RBH). However the values of
$(r_h,r_c,v_c)$ are approximately equal to (1.66,~2.1575,~0.707107)
and (1.32,~1.75626,~0.707107) for KS RBH and DDF RBH, respectively.
For $H=H_c=0.170615$ (KS RBH) and $H=H_c=0.232518$ (DDF RBH), the
solution curves pass through the CPs $(r_c,-v_c)$ and $(r_c,v_c)$.
It is observed that the heteroclinic orbit exists in the range
$-v_c<v<v_c$ and also passes through two CPs $(r_c, -v_c)$ and
$(r_c, v_c)$. It can also be seen that the fluid flow is closer to
DDF RBH instead of KS BH. It is mentioned here that the fluid
experiences the particle emission or fluid flow-out when $v>0$ while
fluid accretes for $v<0$. Moreover, Fig. \textbf{1} describes the
following four types of fluid.
\begin{itemize}
\item We observe the supersonic/subsonic flows out of the fluid in the
ranges $v_c<v<1$ and $0<v<v_c$, while subsonic/supersonic accretion
appear in the ranges $-v_c<v<0$ and $-1<v<-v_c$, respectively.
\item We have purely supersonic outflow for $v>v_c$ and purely
supersonic accretion for $v<-v_c$.
\item We have subsonic flow-out followed by subsonic accretion for
$v_c>v>-v_c$.
\item We have supersonic out flow followed by subsonic motion,
(upper plot) and subsonic accretion followed by supersonic accretion
(lower plot).
\end{itemize}
According to above discussion, we can say that the fluid outflow
starts at horizon because of its high pressure which leads to
divergence (can be seen from Eq.(\ref{pp2})) and the fluid under the
effects of its own pressure flows back to spatial infinity
\cite{25}. From Fig. \textbf{1}, we observe that the supersonic
accretion followed by subsonic accretion and ends inside the horizon
which does not give support to the claim that ``the flow must be
supersonic at the horizon" \cite{38}. Thus, the flow of the fluid is
neither transonic nor supersonic near the horizon \cite{39,40}.
These new solutions correspond to fine tunning and instability
problems in dynamical systems.The issue of stability is related to
the nature of saddle points (CPs ($r_c,v_c$) and ($r_c,-v_c$)) of
the Hamiltonian function. Further analysis of stability could be
done by using Lyapunov's theorem or linearization of dynamical
system \cite{41,42,43} and their variations \cite{44}. Another
stability issue is the outflow of the fluid starts in the
surrounding of horizon under the effect of pressure divergent. This
outflow is unstable because it follow a subsonic path passing
through the saddle point ($r_c, v_c$) and becomes supersonic with
speed approaches the speed of the light. The point ($r=r_h, v=0$)
can be observed as attractor as well as repeller where solution
curves converge and diverge, respectively, in the cosmological point
of view \cite{25,44}.
\begin{figure}
\centering
\begin{tabular}{@{}cccc@{}}
\includegraphics[width=.5\textwidth]{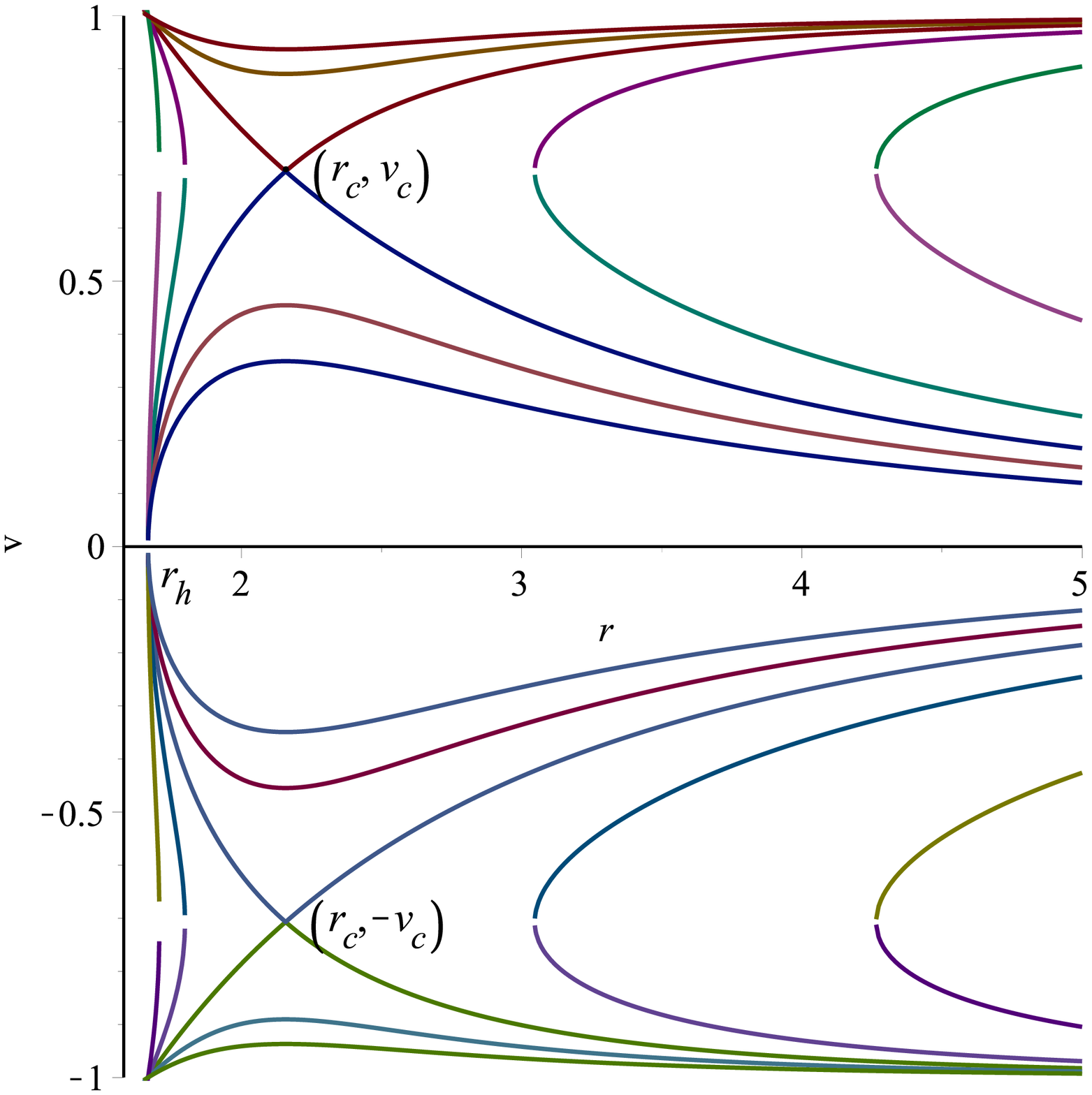} &
\includegraphics[width=.5\textwidth]{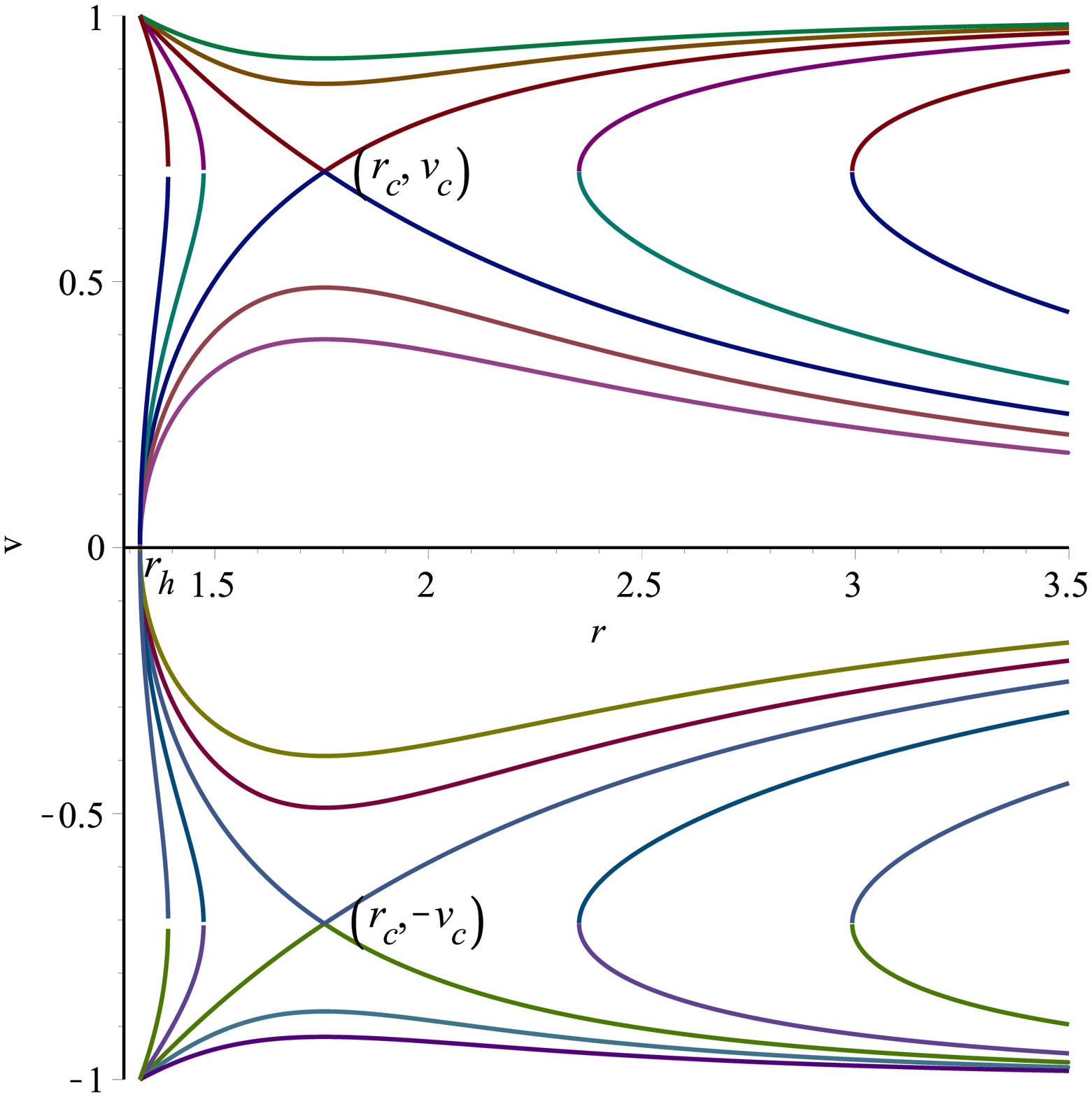}
\end{tabular}
\caption{Case $k=1/2$. Left Panel: Plot of (\ref{2}) for KS BH with
$M=1$, $b=0.9$. Right Panel: Plot of (\ref{2}) for DDF RBH with
$M=1$, $q=0.6$. }
\end{figure}

\subsection{Solutions for $(k=1/3)$
and $(k=1/4)$: Separatrix heteroclinic flows}

The EoS with $k=1/3$ and $k=1/4$ describes the fluid such as photon
gas and Sub-relativistic fluids (those fluids whose energy density
exceeds their isotropic pressure), respectively. The Hamiltonian
(\ref{is5}) for above mentioned fluids takes the following form
\begin{eqnarray}\label{k1} H &=& \frac{X^{2/3}(r)}{r^{4/3}
|v|^{2/3} (1-v^2)^{2/3}},~~~~~~~~ \text{for}~~~~k=1/3\\\label{k2} H
&=& \frac{X^{3/4}(r)}{r |v|^{1/2} (1-v^2)^{3/4}},~~~~~~~~~~
\text{for}~~~~k=1/4.
\end{eqnarray}

\begin{figure}
\centering
  \begin{tabular}{@{}cccc@{}}
    \includegraphics[width=.5\textwidth]{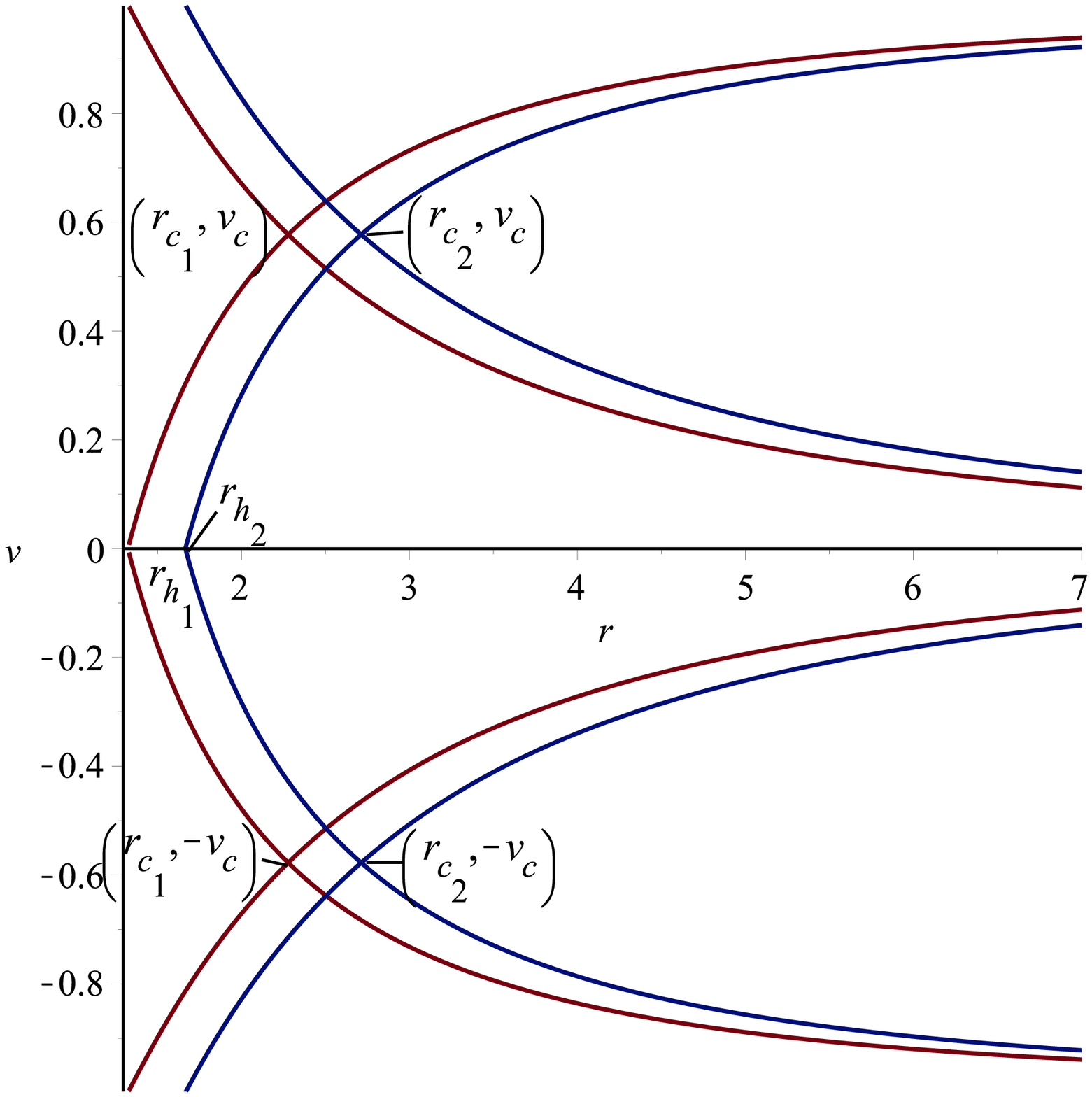} &
    \includegraphics[width=.5\textwidth]{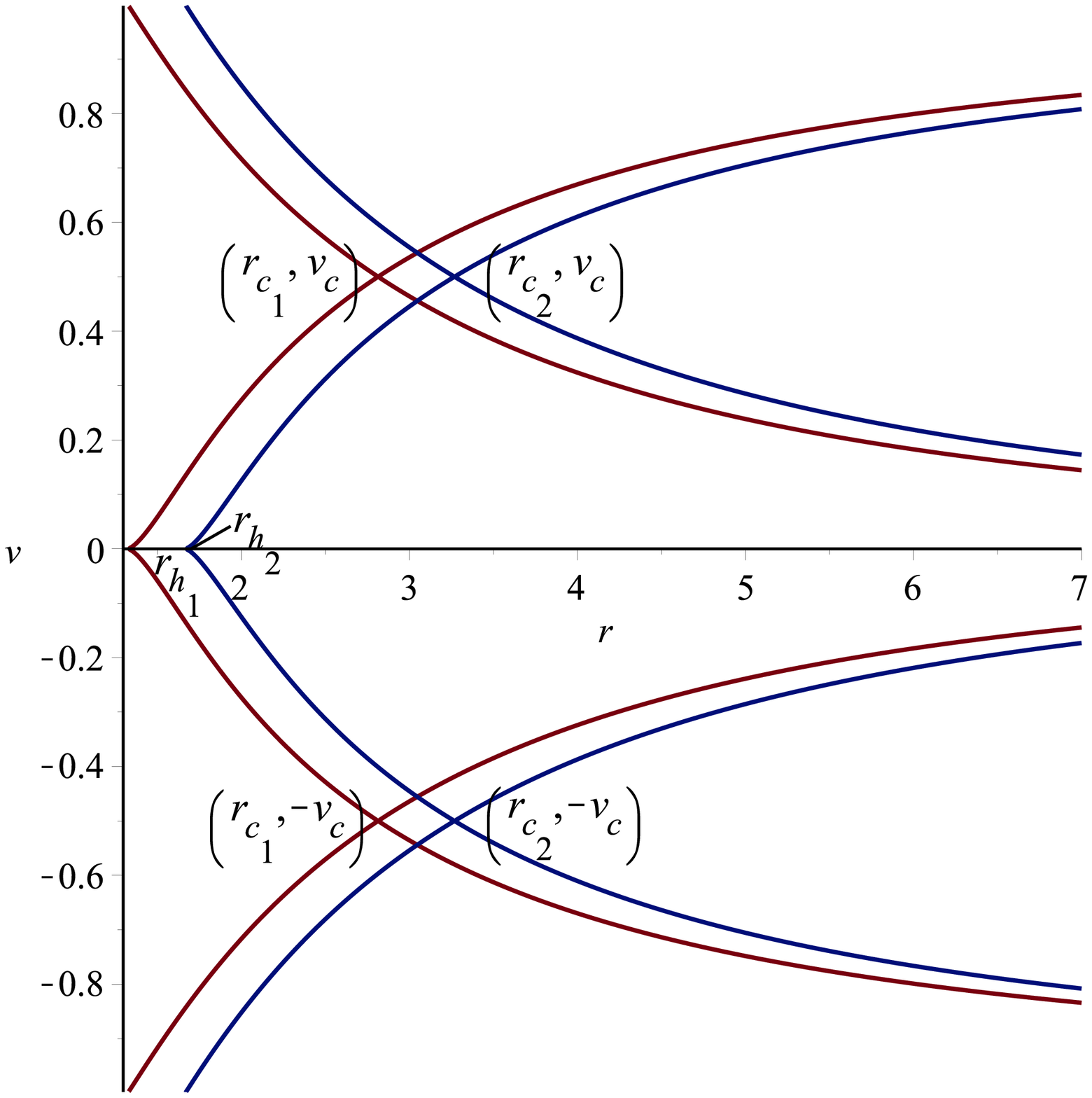}
    \end{tabular}
    \caption{Left panel: Case $k=1/3$. Red plot of (\ref{k1})
    for DDF RBH with $q=0.6$ and $M=1$. The parameters are
    $r_h \simeq 1.33$, $r_{c_1}=2.2787$ and $v_c=\frac{1}{\sqrt{3}}
    \simeq0.57736$. The solution curve passes through the saddle
    CPs $(r_{c_1},v_c)$ and $(r_{c_1},-v_c)$ for which
    $H=H_c\simeq0.261266$. Blue plot of (\ref{k1})
    for KS RBH with $b=0.9$ and $M=1$. The parameters are
    $r_h \simeq 1.68$, $r_{c_2}=2.71329$ and $v_c=\frac{1}{\sqrt{3}}
    \simeq0.57736$. The solution curve passes through the saddle
    CPs $(r_{c_2},v_c)$ and $(r_{c_2},-v_c)$ for which
    $H=H_c\simeq0.22373$. Right panel: Case $k\leq1/4$. Red plot of (\ref{k2})
    for DDF RBH with $q=0.6$ and $M=1$. The parameters are
    $r_h \simeq 1.367$, $r_{c_1}=2.81368$ and $v_c=0.5$. The solution curve passes through the saddle
    CPs $(r_{c_1},v_c)$ and $(r_{c_1},-v_c)$ for which
    $H=H_c\simeq0.29985$. Blue plot of (\ref{k2})
    for KS RBH with $b=0.9$ and $M=1$. The parameters are
    $r_h \simeq 1.717$, $r_{c_2}=3.2682$ and $v_c=0.5$. The solution curve passes through the saddle
    CPs $(r_{c_2},v_c)$ and $(r_{c_2},-v_c)$ for which
    $H=H_c\simeq0.273276$.}
  \end{figure}

\begin{figure}
\centering
  \begin{tabular}{@{}cccc@{}}
    \includegraphics[width=.5\textwidth]{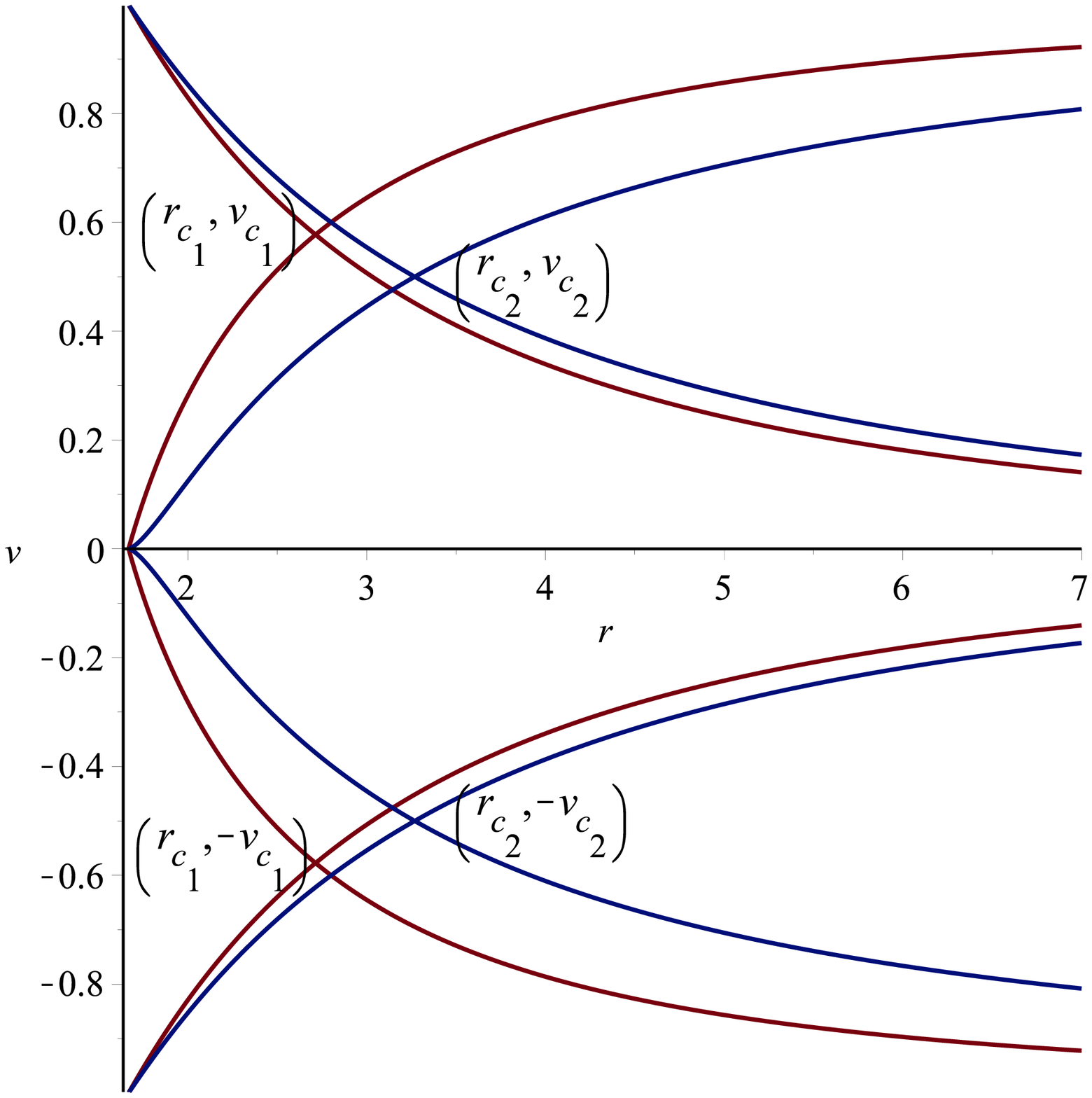} &
    \includegraphics[width=.5\textwidth]{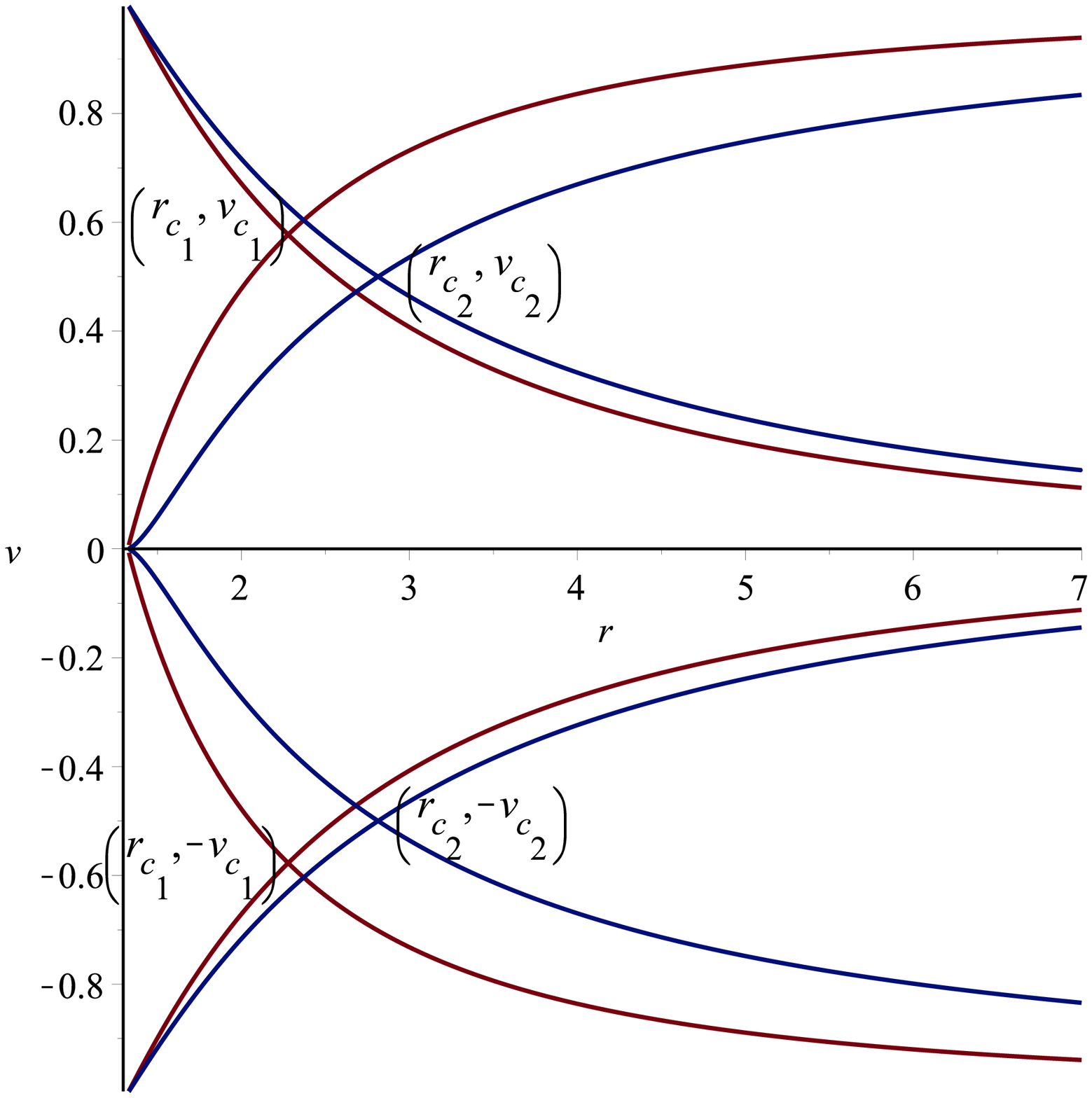}
    \end{tabular}
    \caption{Case $k\leq1/3$. Left panel: Red plot of (\ref{k1}) for KS RBH with $b=0.9$
     and $M=1$ ($k=1/3$). The parameters are $r_h \simeq 1.717$, $r_{c_1}=2.71329$ and $v_{c_1}=\frac{1}{\sqrt{3}}
    \simeq0.57736$. Blue plot of (\ref{k2}) for KS RBH $b=0.9$
     and $M=1$ ($k=1/4$). The parameters are $r_h \simeq 1.717$, $r_{c_2}=3.2682$ and $v_{c_2}=0.5$. Right panel:
     Red plot of (\ref{k1}) for DDF RBH with $q=0.6$
     and $M=1$ ($k=1/3$). The parameters are $r_h \simeq 1.33$, $r_{c_1}=2.2787$ and $v_{c_1}=\frac{1}{\sqrt{3}}
    \simeq0.57736$. Blue plot of (\ref{k2}) for DDF RBH $b=0.9$
     and $M=1$ ($k=1/4$). The parameters are $r_h \simeq 1.33$, $r_{c_2}=2.2787$ and $v_{c_2}=0.5$.}
  \end{figure}

It is clear from (\ref{k1}) and (\ref{k2}) that $(r, v^2)=(r_h, 1)$
are not CPs of dynamical system. Figure \textbf{2} represents the
contour plots of (\ref{k1}) and (\ref{k2}) in $(r, v)$ for $k=1/3$
(left panel) and $k=1/4$ (right panel). We compare the fluid flow
for DDF RBH (red plot) and KS RBH (blue plot) in left panel
($k=1/3$) and right panel ($k=1/4$). For the case $k=1/3$, there are
two saddle points ($r_{c_1}, -v_{c_1}$) and ($r_{c_1}, v_{c_1}$) for
DDF RBH and two saddle points ($r_{c_2}, -v_{c_2}$) and ($r_{c_2},
v_{c_2}$) for KS RBH. It is observed that the subsonic flow proceeds
from upper branch in lower plot, passes through CPs ($r_{c_1},
-v_{c_1}$) (DDF RBH) and ($r_{c_2}, -v_{c_2 }$) (KS RBH), then
crosses the horizon $r_{h_1}$ (DDF RBH) and $r_{h_2}$ (KS RBH). On
the other hand, supersonic flows proceeds from the lower branch in
lower plot again passes through the CPs ($r_{c_1}, -v_{c_1}$) and
($r_{c_2}, -v_{c_2}$); as the fluid approaches the horizon then $v$
vanishes. The supersonic accretion proceeds from the upper branch of
upper plot and passes through CPs ($r_{c_1}, v_{c_1}$) (DDF RBH) and
($r_{c_2}, v_{c_2}$) (KS RBH) and so on. For subsonic accretion,
fluid proceeds from the lower branch of upper plot and passes
through CPs and crosses the horizons. Similar behavior of fluid
could be observed for the case $k=1/4$ in the right panel of Figure
\textbf{2}.

The left panel of Figure \textbf{3} represents the comparison of
(\ref{k1}) ($k=1/3$) and (\ref{k2}) ($k=1/4$) for KS RBH. It can be
seen that if the value of $k$ increases, then the saddle points
shifted towards the KS RBH. Also, the right Panel of Figure
\textbf{3} represents the comparison of (\ref{k1}) ($k=1/3$) and
(\ref{k2}) ($k=1/4$) for DDF RBH. Similar behavior of CPs is
observed for DDF RBH.

\section{ Polytropic test fluids }

The polytropic EoS is
\begin{equation}\label{p1}
    p = G(n) = K n^{\alpha},
\end{equation}
where \emph{K} and $\alpha$ are constants. One can apply the
constraint $\alpha>1$ for ordinary matter. From Eqs. (\ref{a16}) and
(\ref{p1}), one can easily construct the expression of specific
enthalpy:
\begin{equation}\label{p2}
h = m+\frac{K \alpha n^{\alpha -1}}{\alpha-1},
\end{equation}
where \emph{m} is the baryonic mass. Using above equation and the
three dimensional speed of sound (\ref{a18}), we have
\begin{equation}\label{p3}
a^2=\frac{(\alpha -1)U}{m(\alpha-1)+U},
\end{equation}
where $(U \equiv K \alpha n^{\alpha -1})$. Equations (\ref{p2}) and
(\ref{p3}) leads to
\begin{equation}\label{p4}
h = m\frac{\alpha -1}{\alpha - 1 - a^2}.
\end{equation}
Furthermore, using (\ref{h2}) and (\ref{p2}), we have
\begin{equation}\label{p5}
h = m\left(1+Q\left(\frac{1-v^2}{r^4X(r)
v^2}\right)^{(\alpha-1)/2}\right),
\end{equation}
where
\begin{equation}\label{p6}
Q\equiv\frac{K \alpha n_c^{\alpha
-1}}{m(\alpha-1)}\left(\frac{r_{c}^5X(r_c)_{,r_c}}{4}\right)^{(\frac{\alpha-1}{2})}=constant.
\end{equation}
Inserting (\ref{p5}) into (\ref{h2}), the Hamiltonian system takes
the following form
\begin{equation}\label{p7}
    H=\frac{X(r)}{1-v^2}\left(1+Q\left(\frac{1-v^2}{r^4 X(r)
    v^2}\right)^{(\frac{\alpha-1}{2})}\right)^2,
\end{equation}
where $m^2$ has been absorbed into a redefinition of $(\bar{t},H)$.
It is observed that $\frac{dX(r)}{dr}>0$ for all \emph{r} which
implies that $Q>0$ with $\gamma>1$. Hence, the square term in Eq.
(\ref{p7}) is positive while the coefficient $\frac{X(r)}{1-v^2}$
diverges as \emph{r} approaches to infinity $(0\leq1-v^2<1)$. Thus
the Hamiltonian also diverges.
\begin{figure}
\centering
  \begin{tabular}{@{}cccc@{}}
    \includegraphics[width=.5\textwidth]{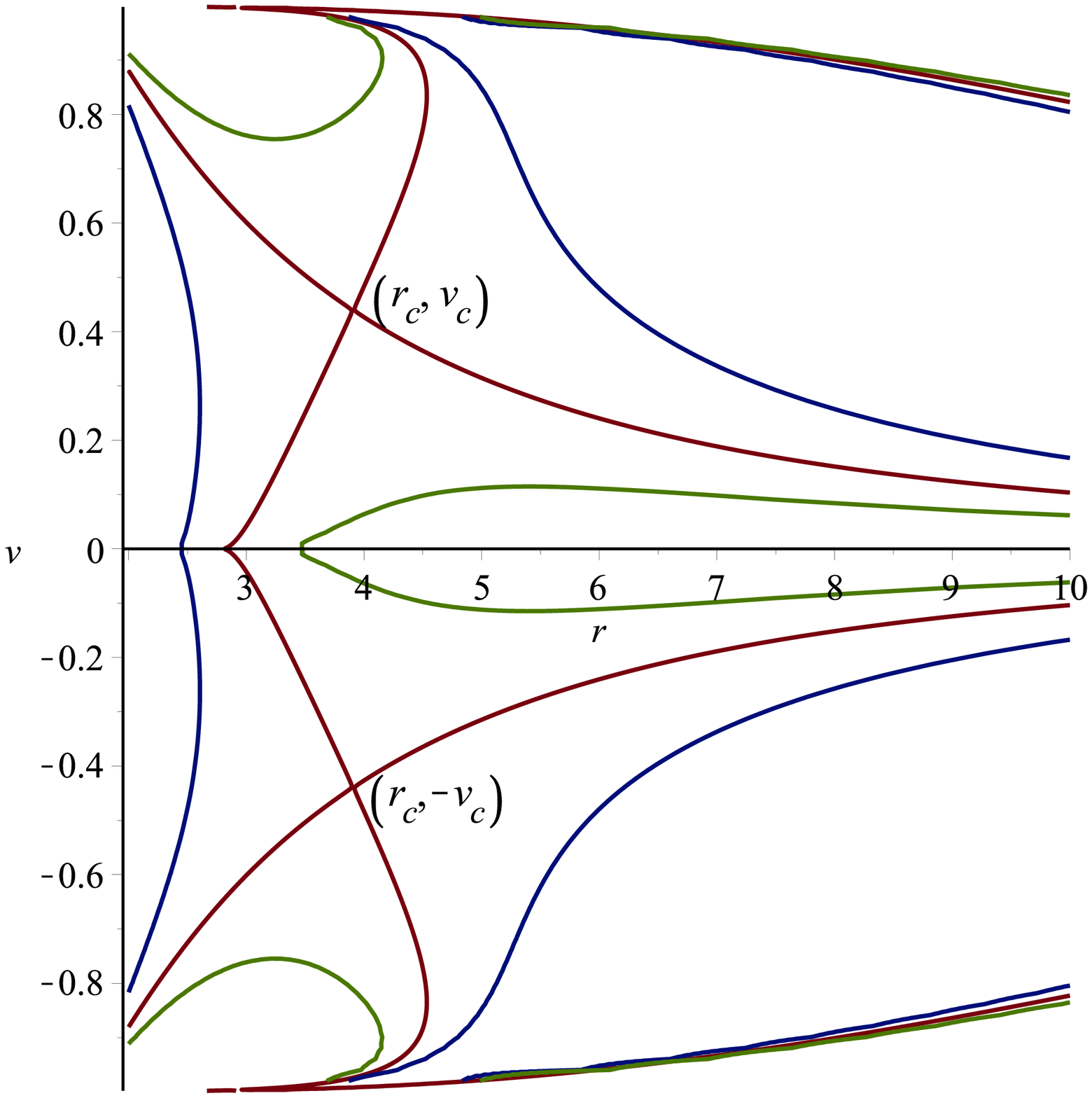} &
    \includegraphics[width=.5\textwidth]{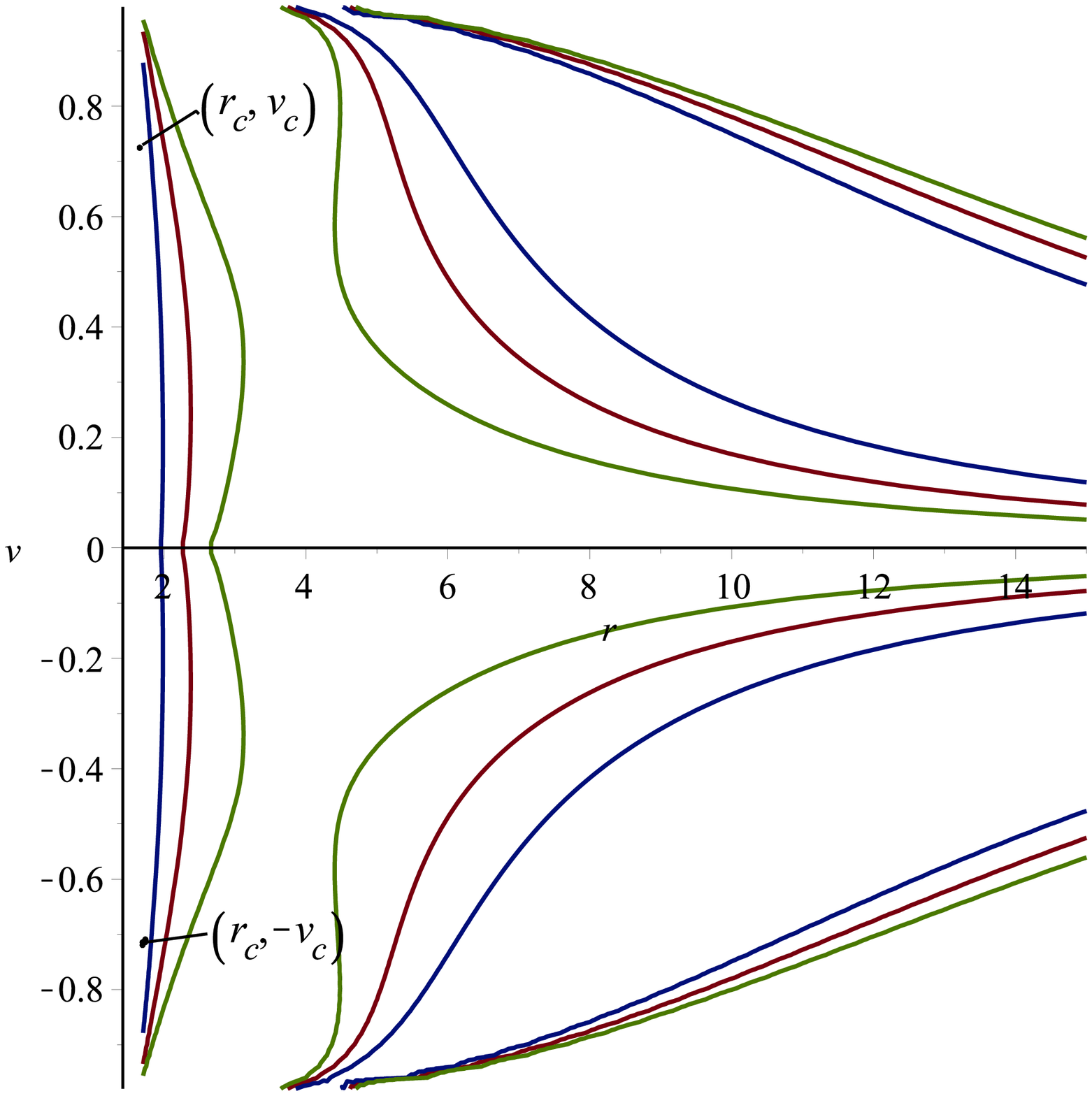}
    \end{tabular}
    \caption{Left Panel: Contour plot of (\ref{p7}) for KS RBH with $b=0.9$, $M=1$,
     $\alpha=0.5$, $Y=-1/8$ and $n_c=0.19$. The parameters are $r_h\simeq1.666666667$,
     $r_c\simeq 3.928297284$ and $v_c\simeq 0.4377733181$.
     The Solution curve passing the the CP $(r_c, v_c)$ for $H=H_c\simeq0.3173998409$.
     Right Panel: Contour plot of (\ref{p7}) for DDF BH with $q=0.5$, $M=1$,
     $\alpha=0.5$, $Y=-1/8$ and $n_c=0.19$. The parameters are $r_h\simeq1.614278177$,
     $r_c\simeq 1.988815258$ and $v_c\simeq0.7359530796$.
     The Solution curve avoiding the CP $(r_c, v_c)$ for $H=H_c\simeq 0.2143193585$.}
\end{figure}
Using the technique of authors \cite{24,25}, we finally come to the
following system
\begin{eqnarray}\label{p8}
(\alpha - 1 -v_c^2)\left(\frac{1-v_c^2}{r_c^4 X(r_c)
v_c^2}\right)^{(\alpha -1)/2} &=& \frac{n_c}{2Y}\left(r_c^5
X(r_c)_{r_c}\right)^{\frac{1}{2}} v_c^2,\\\label{p9}
    v_c^2 &=& \frac{r_c X(r_c)_{r_c}}{r_c X(r_c)_{r_c} + 4 X(r_c)}.
\end{eqnarray}
By putting Eq.(\ref{p9}) in (\ref{p8}), we find $r_c$ and then
corresponding $v_c$ from (\ref{p9}). It is observed from
Eq.(\ref{p8}) that $\alpha<1$, $v_c^2>\alpha -1$ as mentioned in
\cite{24}.
\begin{figure}
\centering
  \begin{tabular}{@{}cccc@{}}
    \includegraphics[width=.5\textwidth]{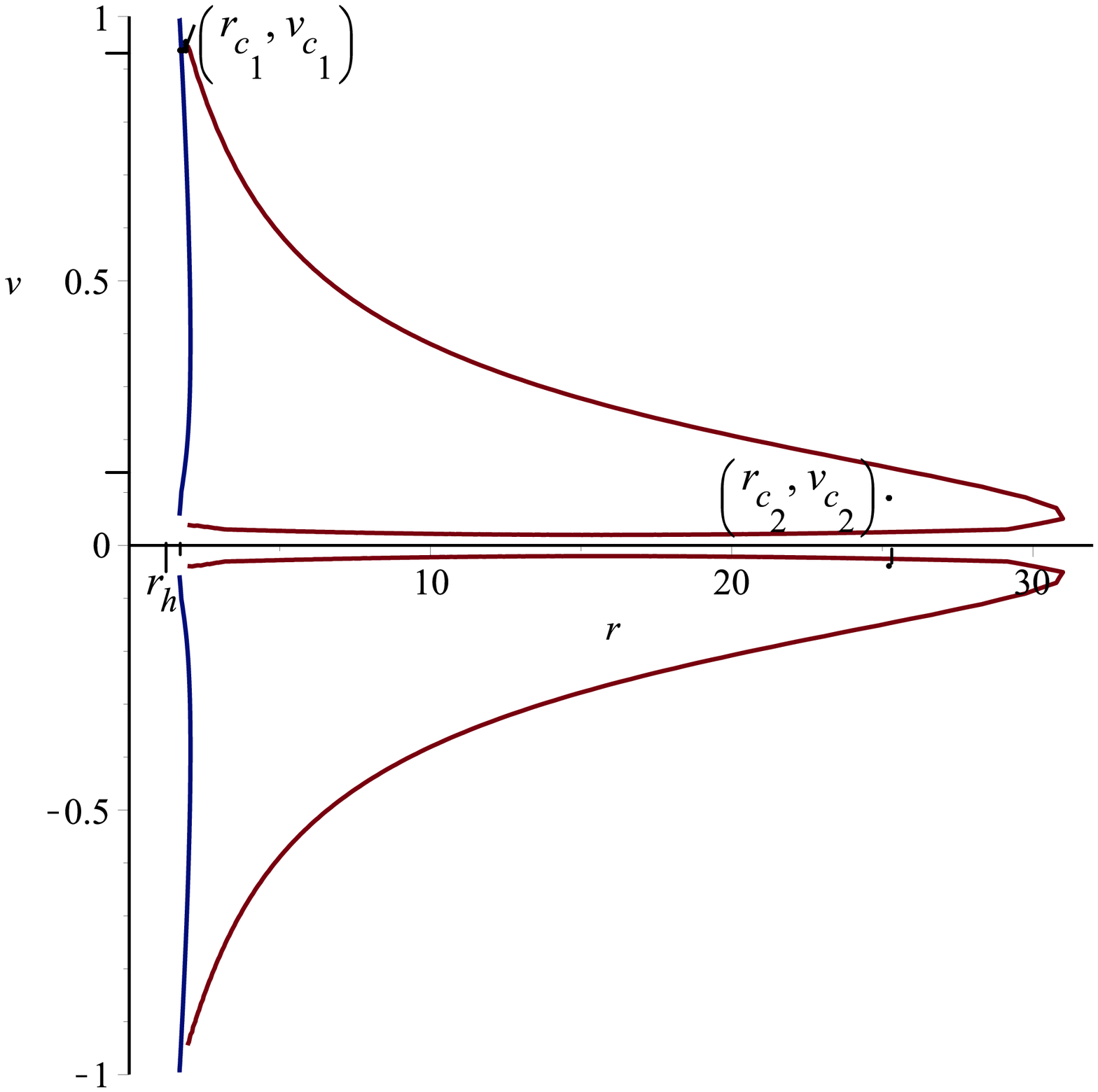} &
    \includegraphics[width=.5\textwidth]{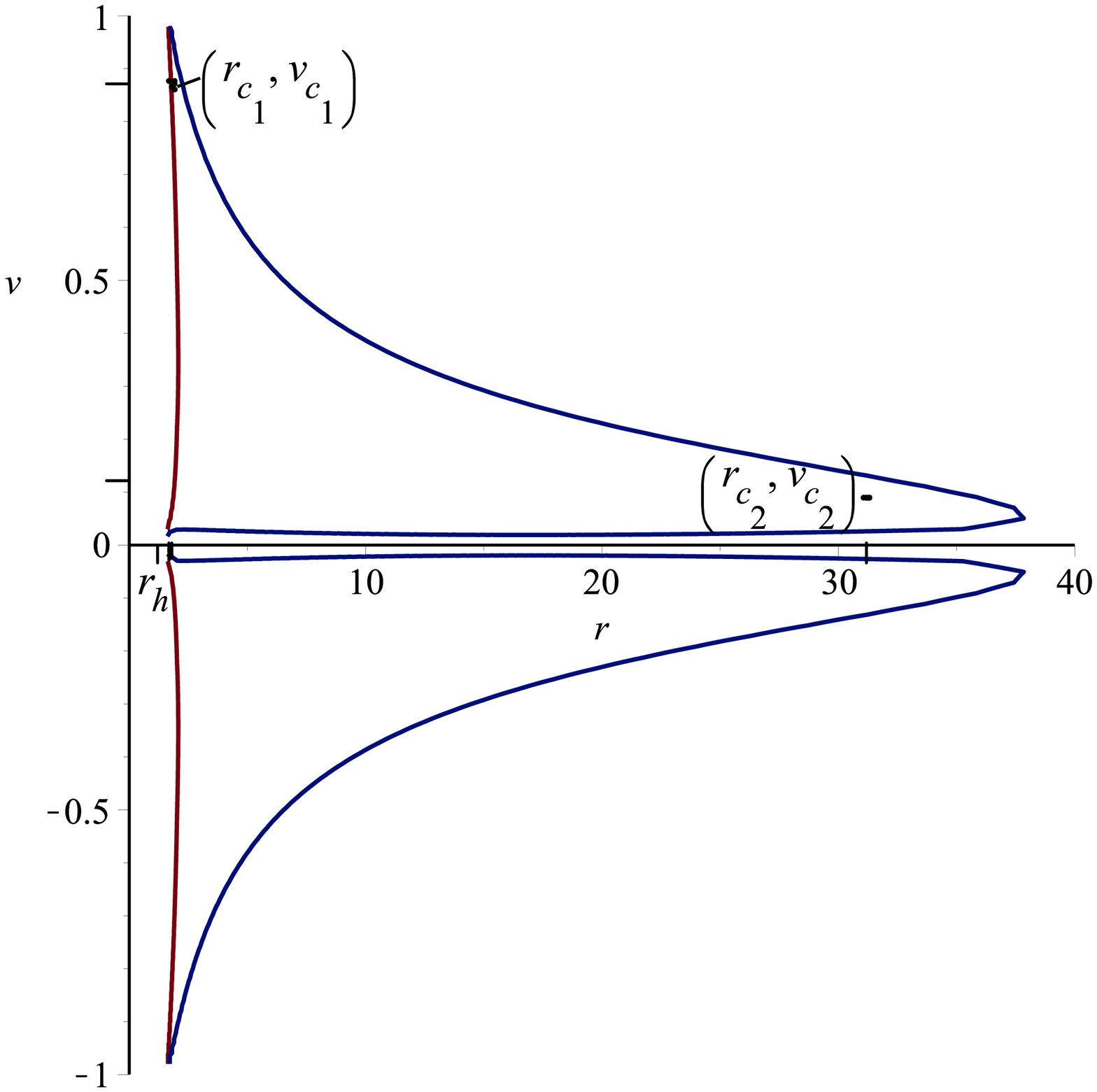}
    \end{tabular}
    \caption{Left Panel: Contour plot of (\ref{p7}) for KS RBH with $b=0.9$, $M=1$,
     $\alpha=1.9$, $Y=1/8$ and $n_c=0.001$. The parameters are $r_h\simeq1.666666667$ , $r_{c_1}\simeq 1.719016508$, $v_{c_1}\simeq 0.9439334159$, $r_{c_2}\simeq 25.49303509$ and $v_{c_2}\simeq
     0.1443387155$. Red plot is the solution curve passing through
     ($r_{c_2}, v_{c_2}$)and ($r_{c_2}, -v_{c_2}$) with
     $H=H_{c_2}=0.9452241039$. Blue plot is the solution curve passing through
     ($r_{c_1}, v_{c_1}$)and ($r_{c_1}, -v_{c_1}$) with
     $H=H_{c_2}=.2028972112$.
     Right Panel: Contour plot of (\ref{p7}) for DDF RBH with $q=0.5$, $M=1$,
     $\alpha=1.8$, $Y=1/8$ and $n_c=0.001$.The parameters are $r_h\simeq1.614278177$ , $r_{c_1}\simeq 1.724674282$, $v_{c_1}\simeq 0.8893889966$, $r_{c_2}\simeq 31.59490067$ and $v_{c_2}\simeq
     0.1283035425$. Blue plot is the solution curve passing through
     ($r_{c_2}, v_{c_2}$)and ($r_{c_2}, -v_{c_2}$) with
     $H=H_{c_2}=0.9576657920$. Red plot is the solution curve passing through
     ($r_{c_1}, v_{c_1}$)and ($r_{c_1}, -v_{c_1}$) with
     $H=H_{c_2}=0.2585395967$.}
\end{figure}
The left panel of Figure \textbf{4} represents the contour plot of
(\ref{p7}) for KS BH with $b=0.9$, $M=1$, $\alpha =0.5095$, $Y= -
1/8$ and $n_c=0.19$. We find the new behavior of the fluid by
finding the exact CP $r_c\simeq 3.928297284$, $v_c\simeq
0.4377733181$ for which $H=H_c\simeq0.3173998409$. It can be
observed that the accretion starts from subsonic flow for
$r\rightarrow\infty$ then follow supersonic passing through the
saddle point and ends into the horizon. However, the supersonic flow
out observed in the vicinity of horizon passing through the saddle
point and ends subsonically for $r\rightarrow\infty$. The right
panel of Figure \textbf{4} represents the contour plot of (\ref{p7})
for DDF RBH with $q=0.5$, $M=1$, $\alpha =0.5$, $Y= - 1/8$ and
$n_c=0.19$. We find the CP at $r_c\simeq 1.988815258$,
$v_c\simeq0.7359530796$ for which $H=H_c\simeq 0.2143193585$. We
notice that the accretion starts from subsonic flow for
$r\rightarrow\infty$ then follow supersonic avoiding the saddle
point and ends into the horizon. However, the supersonic flow out
observed in the vicinity of horizon avoiding the saddle point and
ends subsonically for $r\rightarrow\infty$. These features agrees
with GR BHs \cite{24,25}.

In Fig. \textbf{5}, we are assuming $\alpha = 1.9, Y=1/8, n_c=0.001$
in left panel for KS RBH and $\alpha = 1.8, Y=1/8, n_c=0.001$ in
right panel for DDF RBH which lead to four CPs for each RBH, but
none of them is saddle point. There are three types of flow: $(1)$
non heteroclinic because the solution curves avoiding the CPs and
accretion starting from leftmost point until the horizon, $(2)$
followed by non relativistic flow out and $(3)$ subsonic non-global.
There are two other fluids flow: $(1)$ partly subsonic accretion and
flow-out with source-sink at rightmost point of both graphs and
$(2)$ partly supersonic accretion and flow-out with source-sink at
rightmost point of both graphs. Other solutions could also be found
at $\alpha=1.8$ for KS RBH and at $\alpha=5.5/3$ for DDF RBH. Since
the fluid is considered as a test matter in geometry of BH, we
observe no homoclinic flow i.e. flow following closed path
\cite{24,25}.

\section{Conclusion}

In this work, we studied the steady state spherically symmetric
accretion onto RBHs. Applying the technique of \cite{24,25} and
develop the Hamiltonian dynamical system for tackling the accretion
problem. As the pressure, baryon number density and other densities
which diverge on horizon, while three velocity have remained bounded
by $-1$ and $1$ and hence it does not diverge on horizon. We have
studied the motion of isothermal relativistic, ultra-relativistic,
radiation and sub-relativistic test fluids in the frame work of
Hamiltonian dynamical system around RBHs. The thermodynamical
properties of fluids have been studied for different choices of EoS
parameter. Moreover, CPs and conserved quantities have been found
for different fluids. The behavior of accreting fluid have been
discussed as subsonic and supersonic according to EoS and RBHs. We
compared the fluid flow for two different RBHs and observed that the
fluid flow and CPs are closer to DDF RBH instead of KS RBH (Figure
\textbf{1}). If CP is the saddle point then the solution curve
divides the $(r, v)$ plane into the regions where fluid flow is
physical for higher values of Hamiltonian and unphysical in lower
values of Hamiltonian.

It has been observed that the supersonic accretion followed by
subsonic accretion ends inside the horizon and it does not give
support to the claim that "the flow must be supersonic at the
horizon" \cite{38}. Hence the flow of the fluid is neither transonic
nor supersonic near the horizon \cite{39,40}. These new solutions
correspond to fine tunning and instability problems in dynamical
systems.
Furthermore, we have observed from Figure \textbf{3} that if the
value of EoS parameter \emph{(k)} increases then CPs shifted
\textbf{towards both RBHs}. Hence, it is concluded that the three
velocity depends on CPs and EoS parameter on phase space. It is
interesting to mention here that the results obtain in Figure
\textbf{1} agrees with \cite{24,25}. We have also observed from
right panel of Figure \textbf{4} that accretion starts from subsonic
and ends at supersonic into horizon but avoiding the saddle points
for DDF RBH. Hence no saddle point occurs in the solution of DDF
RBH. Furthermore, the subsonic flow appears to be almost
non-relativistic. These features agrees with GR BHs \cite{24,25}. On
the other hand, in left panel of Figure \textbf{4}, it is seen that
accretion starts from subsonic and ends at supersonic into horizon
but passing through the saddle points for KS RBH. Hence, saddle
point occurs in the solution of KS RBH. These features are different
from \cite{24,25}.

\begin{flushleft}
\textbf{Acknowledgment}
\end{flushleft}
We are thankful to the anonymous referees for the constructive
remarks on our manuscript which have improved our manuscript. We are
also thankful to Prof. Olivier Sarbach for useful discussions and
remarks on our manuscript.

\end{document}